\documentclass[prl,superscriptaddress,showpacs,preprint]{revtex4}
\usepackage{graphicx}% Include figure files
\usepackage{dcolumn}% Align table columns on decimal point
\usepackage{bm}% bold math

\renewcommand{\vec}{\bm}

\begin{document}

\preprint{\today}
\title{Direct observation of electric-quadrupolar order in UO$_{2}$}
\author{S. B. Wilkins}
\affiliation{European Commission, Joint Research Centre, Institute for
   Transuranium Elements, Postfach 2340, Karlsruhe, D-76125 Germany\\}
\affiliation{European Synchrotron Radiation Facility,
   Bo\^\i te Postal 220, F-38043 Grenoble CEDEX, France\\}
\author{R. Caciuffo}
\affiliation{Dipartimento di Fisica ed Ingegneria dei Materiali e del 
Territorio, Universit\`{a} Politecnica delle Marche, I-60131 Ancona, 
Italy\\}
\author{C. Detlefs}
\affiliation{European Synchrotron Radiation Facility,
   Bo\^\i te Postal 220, F-38043 Grenoble CEDEX, France\\}
\author{J. Rebizant}
\author{E. Colineau}
\author{F. Wastin} 
\affiliation{European Commission, Joint Research Centre, Institute for
   Transuranium Elements, Postfach 2340, Karlsruhe, D-76125 Germany\\}
\author{G. H. Lander}
\affiliation{European Commission, Joint Research Centre, Institute for
   Transuranium Elements, Postfach 2340, Karlsruhe, D-76125 Germany\\}

\date{\today}

\begin{abstract}
We report direct experimental evidence for long-range antiferro ordering of the electric-quadrupole moments on the U ions. Resonant x-ray scattering experiments at the uranium $M_{4}$ absorption edge show a characteristic dependence in the integrated intensity upon rotation of the crystal around the scattering vector.  Although quadrupolar order in uranium dioxide was advocated already in the 1960s no experimental evidence for this phenomenon was provided until now. We conclude with a possible model to explain the phase diagram of the solid solutions of UO$_{2}$ and NpO$_{2}$. We suggest that in the region $0.30 < x < 0.75$ neither the transverse nor the longitudinal quadrupole ordering can dominate,  leading to frustration and only short-range ordering. 

\end{abstract}

\maketitle

Uranium dioxide (\emph{fcc} CaF$_{2}$ crystal structure with space group $Fm\overline{3}m$) is the most studied of any actinide material because of readily available single crystals and industrial applications. Since the 1950s, many important experiments have been performed to elucidate the nature of the low-temperature ground state of UO$_{2}$. These include the determination of the magnetic structure below $T_{N}$ (= 31~K) and value of the magnetic moment\cite{willis188,frazer1448}, the strong softening of the $c_{44}$ elastic constant starting at temperatures considerably above $T_{N}$\cite{brandt11,brandt528}, the evidence for strong magnon-phonon coupling\cite{cowley683}, the observation of an internal re-arrangement of the oxygen atoms below $T_{N}$\cite{faber1770,faber1151}, and evidence that the magnetic structure is of the $3\vec{k}$ variety\cite{burlet121}. Further evidence supporting the non-collinear $3\vec{k}$ ordering was provided by neutron inelastic scattering experiments, which determined the energies of the excited crystal-field levels\cite{amoretti1856}, and NMR experiments on the $^{235}$U and $^{17}$O nuclei below $T_{N}$\cite{ikushima104404}.  

In addition to the wealth of experimental data, important theories were developed, notably the work of Allen \cite{allen530,allen492}, Sasaki and Obata \cite{sasaki1157}, Siemann and Cooper \cite{siemann2869}, and Solt and Erd\"os \cite{solt4718}. Even in the last 3 years a number of new theoretical papers have been published\cite{kudin266402,laskowski140408,magnani054405,ippolito064419}. To a lesser or greater extent, all these theories emphasize the importance of the interplay between the Jahn-Teller and quadrupolar interactions in UO$_{2}$. Allen \cite{allen530,allen492} indeed proposed in 1968 that the uranium electric quadrupoles ordered and the subsequent internal strain led to a change in the positions of the oxygen atoms without an external distortion of the lattice, which remains cubic. 

Experimentally, however, no direct evidence for the ordering of electric quadrupole moments below $T_{N}$ has been presented. This Letter provides that key evidence.  Of course, the observation of an internal distortion of the oxygen atoms at $T_{N}$\cite{faber1770,faber1151} gives indirect proof that quadrupolar effects are important, and the presence of quadrupoles at the U site is strongly inferred from the $^{235}$U NMR results\cite{ikushima104404}. By using resonant x-ray scattering at the U $M_{4}$ resonance, where we probe the $5f$ valence states of UO$_{2}$, we show conclusively the long-range nature of the $5f$ electric quadrupolar ordering. We further show that temperature dependence of this quadrupolar ordering is similar to that associated with the internal distortion of the oxygen cage (as expected), and demonstrate the difference between the electric quadrupolar ordering found in UO$_{2}$ and that reported for NpO$_{2}$\cite{nikolaev054112}. We conclude by recalling studies of the mixed oxide systems (U$_{1-x}$Np$_{x}$)O$_{2}$ and speculate on the possible role of  ``quadrupolar frustration'' in explaining some of the unusual features of this phase diagram. 

Experiments were carried out on the magnetic scattering beamline ID20 at the European Synchrotron Radiation Facility, Grenoble, France. The sample was mounted in a closed cycle refrigerator capable of reaching a base temperature of 12~K. This in turn was mounted within a 5-circle vertical diffractometer. Polarization analysis of the scattered beam was performed using the (111) reflection from a Au analyzer crystal for which the Bragg angle at the M$_{4}$ edge of uranium is close to Brewster's angle.

The resonant X-ray scattering amplitude for an electric dipole (E1) transition can be written in a general form as
\begin{equation}
f^{\mathrm{xres}}_{E1} = f_{0} + if_{1} + f_{2}
\end{equation}
where the terms $f_{n}$ are given by the following equations
\begin{eqnarray}
f_{0} & = &\vec{\epsilon}\cdot\vec{\epsilon}\;\; [F_{11} + F_{1-1}]\nonumber\\
f_{1} & = &(\vec{\epsilon}' \times \vec{\epsilon})\cdot \hat{z}\;\; [F_{11}-F_{1-1}]\nonumber\\
f_{2} & = &\vec{\epsilon} \cdot \tilde{T} \cdot \vec{\epsilon}\;\; [2F_{10} - F_{11} - F_{1-1}].\label{eq:crosssec}
\end{eqnarray}
where $F_{1q}$ is the resonant energy factor, as given by Hannon and Trammell\cite{hannon1248}, and $\vec{\hat{z}}$ is the direction of the magnetic moment. For $\sigma$ incident polarization the terms in $f_{1}$ perform a rotation of the polarization of the incident x-rays. However, terms in $f_{2}$ may or may not result in a rotation of the scattered x-rays. 

The terms in $f_{0}$ do not depend on multipole moments and can be neglected in this work. The term $f_{1}$ probes a tensor of rank 1, with odd time-reversal symmetry arising from a net spin polarization, a difference between overlap integrals, resonant energy, or life-time for the two channels\cite{hill236,blume3615}. The term $f_{2}$ probes a tensor of rank 2, even in time-reversal symmetry. This can arise from an asymmetry intrinsic to the crystal lattice (Templeton or
anisotropic tensor susceptibility scattering\cite{templeton133}) or it can be due to antiferro order of electric-quadrupole moments.

For the case of electric-quadrupole moments, we first evaluate the quadrupolar operator for the $3\vec{k}$ structure of UO$_{2}$ given by
\begin{equation}
  \vec{T}_{ij} = \mu_{i} \mu_{j} 
  - \frac{1}{3}\delta_{ij}\sum_k (\mu_{k} \mu_{k}) 
\end{equation}
where $\vec{\mu}$ is the principal axis of the electric quadrupole moment. From this we can obtain a ``tensorial structure factor'' $\tilde{T}_{Q}$, summing over all atoms within the unit cell. The scattered intensity is then deduced using Eq.~\ref{eq:crosssec} and given by
\begin{equation}
I(\vec{Q}) \propto | \vec{\epsilon}' \cdot \tilde{T}_{\vec{Q}} \cdot \vec{\epsilon} |^{2}
\end{equation}
where $\vec{\epsilon}$ and $\vec{\epsilon}'$ are unit vectors along the incident and scattered x-rays electric field vectors respectively.  For the $3\vec{k}$ transverse structure of UO$_{2}$ there are two S-domains which for scattering vectors of $\vec{Q} = (003)$ and $\vec{Q} = (112)$ yield tensors of the form 
\begin{eqnarray}
\tilde{T}^{A} & = &\left ( \begin{array}{ccc}
0 & 0 & 0 \\ 0 & 0 & 1 \\ 0 & 1 & 0 \\
\end{array}\right ) \label{eqn:domaina}\\
\tilde{T}^{B} & = & \left ( \begin{array}{ccc}
0 & 0 & 1\\ 0 & 0 & 0 \\ 1 & 0 & 0 \\
\end{array}\right ).\label{eqn:domainb}
\end{eqnarray}

The reflections arising from these quadrupoles coincide with those due to magnetic dipole ordering. The experimental challenge is therefore to separate the two. This can be achieved by polarization analysis as for $\sigma$ polarized incident x-rays all scattering from the magnetic dipoles is $\sigma\rightarrow\pi$ with the signal from the electric-quadrupoles being scattered $\sigma\rightarrow\sigma$ \emph{and} $\sigma\rightarrow\pi$. However, previous attempts to observe electric-quadrupolar scattering in UO$_{2}$ have failed for the following reason: It is usual in such systems to work in specular geometry, which for a [001] face crystal allows one to observe scattering at (001) and (003) at the U M$_{4}$ edge. In this configuration quadrupolar scattering from such reflections occurs only in the $\sigma\rightarrow\pi$ channel, and given the strong magnetic dipole scattering at the same wavevector it is therefore impossible to observe. On the other hand, by using an off-specular reflection, such as the (112) used in these measurements, the signal from the electric-quadrupolar scattering is partially $\sigma\rightarrow\sigma$ and therefore observable.

\begin{figure}
\includegraphics[width=\columnwidth]{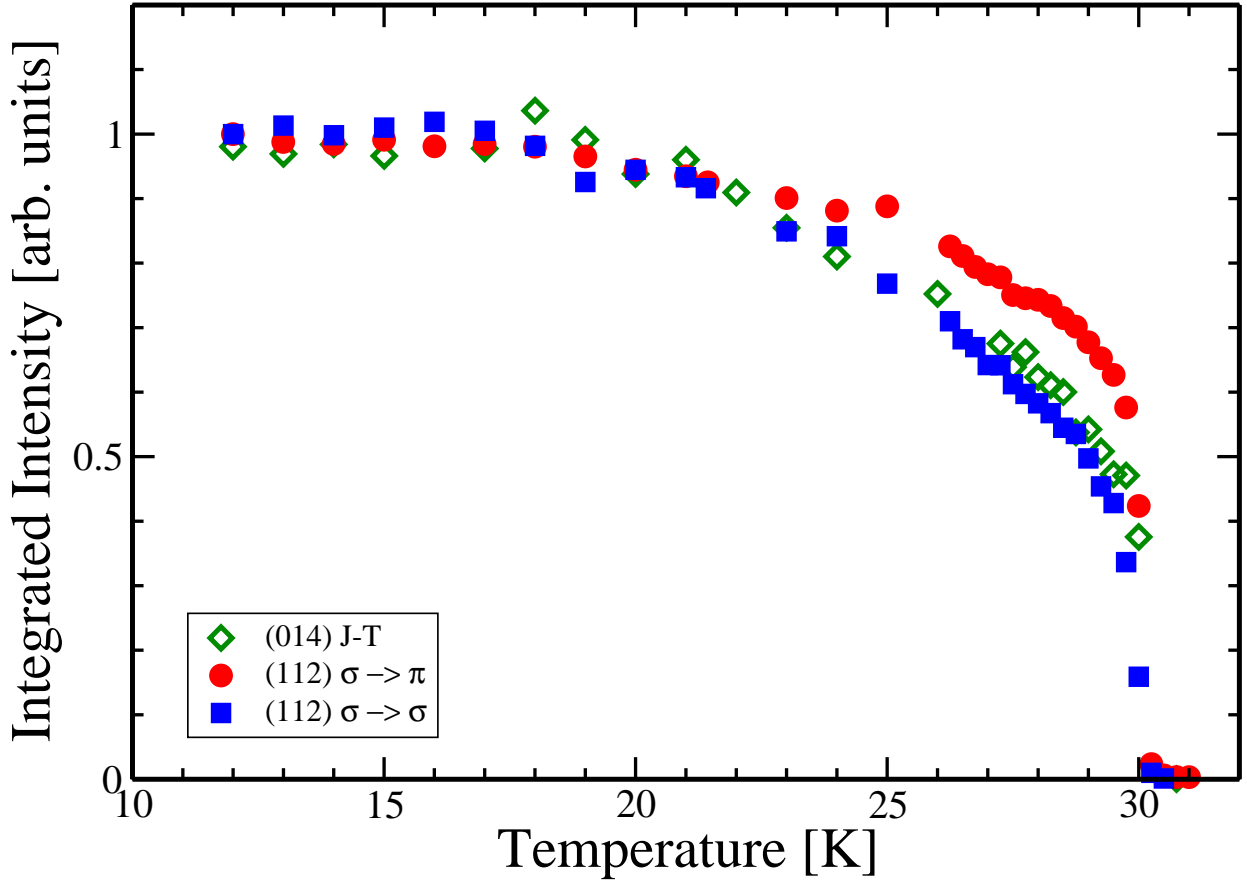}
\caption{Temperature dependence of the integrated intensity of the (112) reflection in both $\sigma \rightarrow \pi$ (circles) and $\sigma \rightarrow \sigma$ (diamonds) polarization channels and the (014) Bragg reflection (measured at 10~keV). These represent the magnetic dipole, electric-quadrupole order and Jahn-Teller distortion respectively.}
\label{fig:tdep}
\end{figure}

Figure~\ref{fig:tdep} shows the integrated intensity as a function of temperature for the (014) and (112) reflections. Scattering from the (014) was collected at an incident photon energy of 10~keV, far away from the resonance condition and corresponds to the internal distortion of the oxygen sub-lattice\cite{faber1770,faber1151}. For the (112) an incident energy of 3.728~keV was used, corresponding to the U $M_{4}$ absorption edge. In this case both $\sigma\rightarrow\pi$ and  $\sigma\rightarrow\sigma$ polarization channels are shown. The former is dominated by the magnetic dipole order parameter while the latter arises solely from the electric quadrupole order. From these data it is apparent that the electric quadrupolar order parameter follows the temperature dependence of the internal distortion. 

\begin{figure}
\includegraphics[width=\columnwidth]{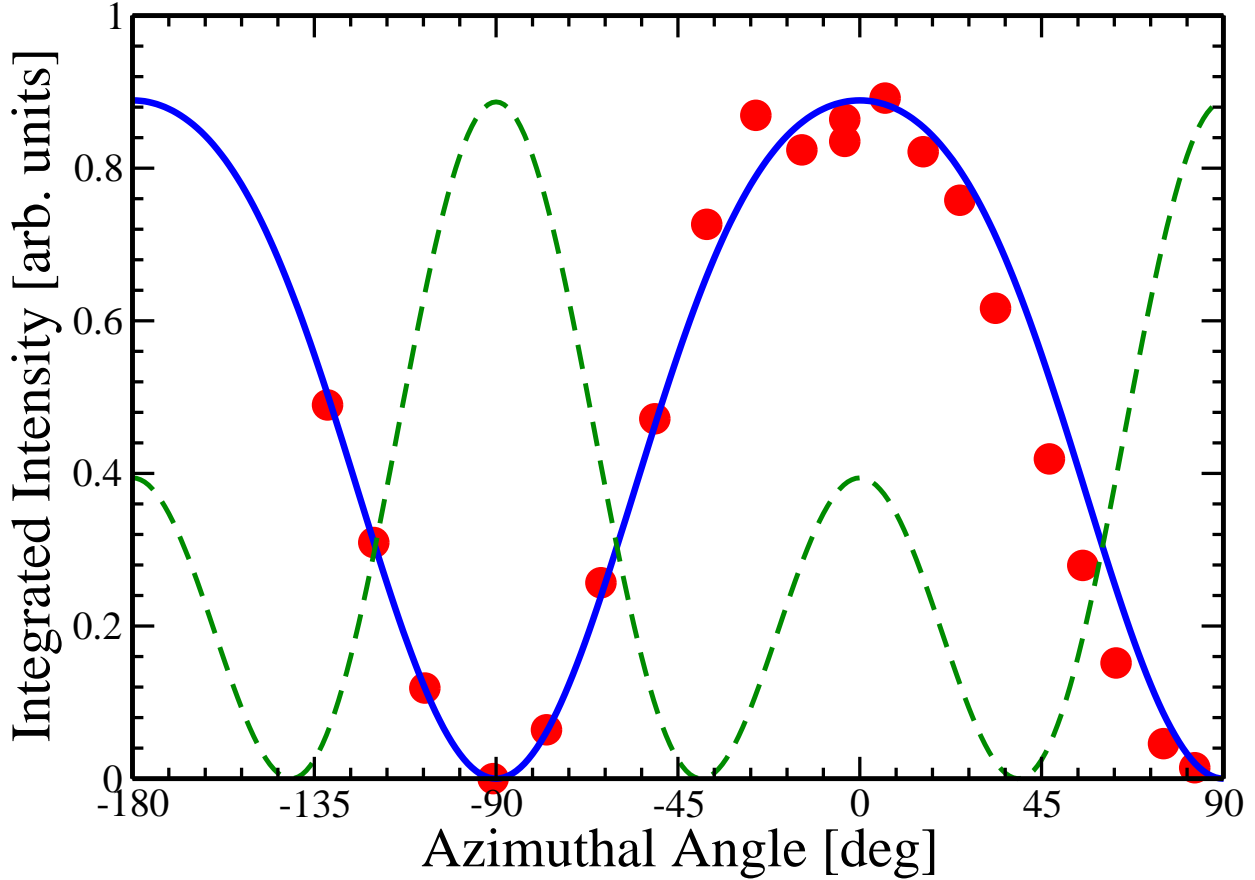}
\caption{The integrated intensity as a function of azimuthal angle $\Psi$ for the (112) reflection measured in the $\sigma \rightarrow \sigma$ polarization channel (Circles). The solid (dashed) line shows the expected azimuthal dependence for the transverse (longitudional) model as described in the text. }
\label{fig:azimuth}
\end{figure}

\begin{figure}
\includegraphics[height=0.65\textheight]{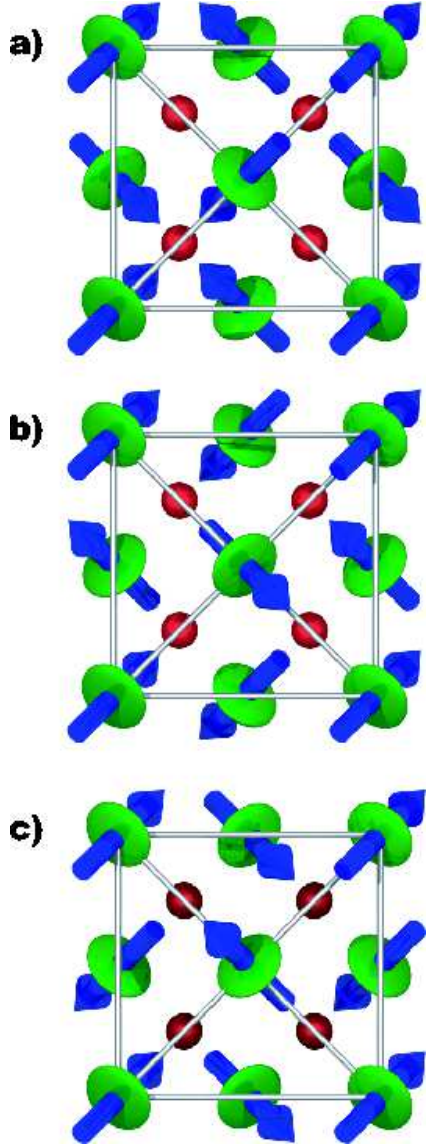}
\caption{Schematic representation of the projection onto the a-b plane of the 3$\vec{k}$ magnetic and electric-quadrupole ordering for the longitudinal (a) configuration and the two S-domains of the transverse configuration (b,c). The magnetic (dipole) moments are represented by blue arrows whereas the electric-quadrupole moments are shown as the green ellipsoids. The red spheres represent oxygen atoms.}
\label{fig:cartoon}
\end{figure}

As the sample is rotated around the scattering vector (azimuthal rotation) the intensity of the superlattice peak exhibits a characteristic oscillation due to the relative rotation of the electric field vector with respect to the crystal lattice. It is this oscillation which enables us to determine the origin of the observed signal. Figure~\ref{fig:azimuth} shows the integrated intensity for the (112) reflection as a function of the azimuthal angle ($\Psi$) in the $\sigma\rightarrow\sigma$ polarization channel. The data were collected at a temperature of 12~K. The origin of $\Psi$ corresponds to the condition when the $[\overline{1}10]$ direction lies within the scattering plane. For this reflection, the azimuthal dependence of the intensity is given by a incoherent addition of the two transverse S-domains (see Fig~\ref{fig:cartoon}),
\begin{equation}
	I_{(112)} = A \left[ |\vec{\epsilon}' \cdot \tilde{T}_{A} \cdot \vec{\epsilon}|^{2} + |\vec{\epsilon}' \cdot \tilde{T}_{B} \cdot \vec{\epsilon}|^{2} \right]
\end{equation}
where $\tilde{T}_{A}$ and $\tilde{T}_{B}$ are the scattering tensors given in equations \ref{eqn:domaina} and \ref{eqn:domainb} respectively. This can be evaluated to give
\begin{equation}
	I_{(112)} = \frac{4}{3}A\cos^{2}\Psi\left[ 1 - \frac{1}{3}\cos^{2}\Psi\right]\label{eqn:azimuth}
\end{equation}
A comparison between Equation~\ref{eqn:azimuth} and the experimental data is shown in Fig~\ref{fig:azimuth}. The dashed line in Fig~\ref{fig:azimuth} corresponds to the case of a longitudinal structure. This shows unambiguously that, with respect to the propagation vector, the orientation of the electric quadrupoles in UO$_{2}$ is {\it transverse} contrary to the case of NpO$_{2}$ in which a {\it longitudinal} orientation is found\cite{paixao187202,caciuffos2287}. Interestingly, because of the crystal-field ground state \cite{magnani1} 
there is \emph{no} magnetic-dipole order in NpO$_{2}$. However, both UO$_{2}$ and NpO$_{2}$ are now established to have electric-quadrupole ordering at low temperature, and in both materials the ordering is $3\vec{k}$ in nature\cite{burlet121,amoretti1856,ikushima104404,paixao187202,caciuffos2287,tokunaga137209}

\begin{figure}
\includegraphics[width=0.9\columnwidth]{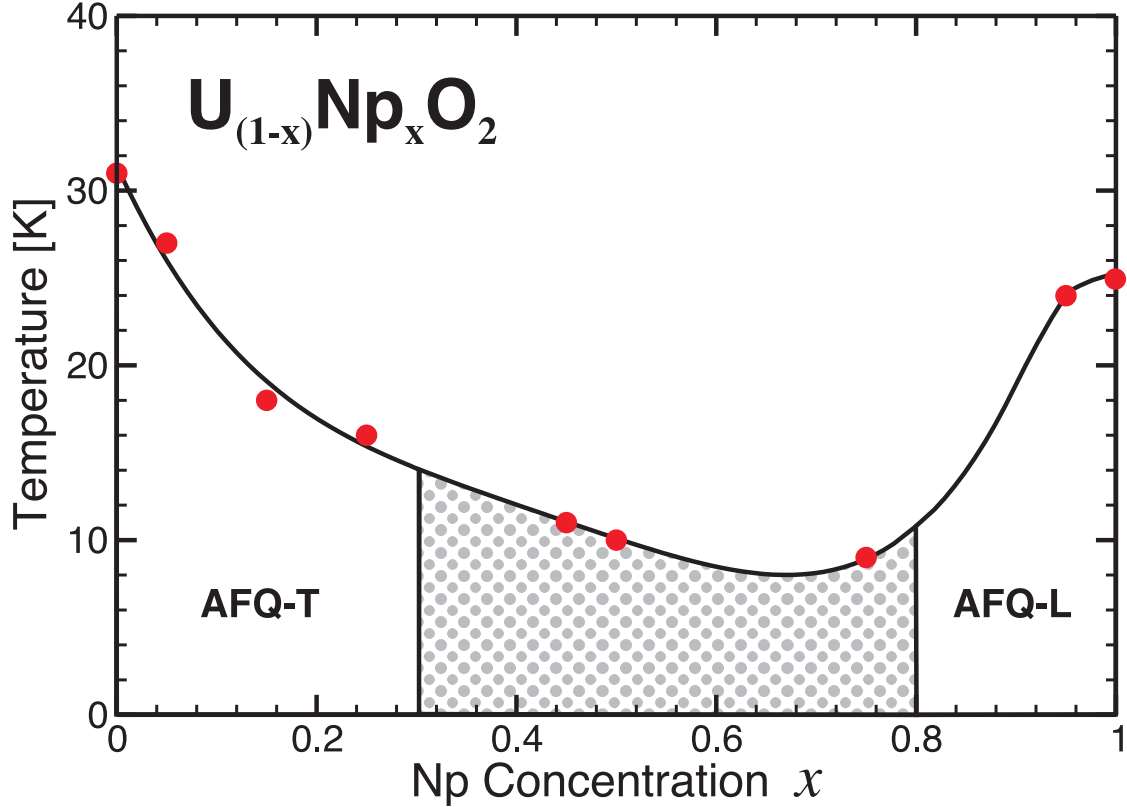}
\caption{Suggested phase diagram for the U$_{(1-x)}$Np$_{x}$O$_{2}$ system as derived from various measurements. The known regions of anti-ferroquadrupolar order are marked by AFQ-T (AFQ-L) for the transverse(longitudional) case. The shaded region in the center denotes the fustrated region of the phase diagram. (see text for details).}
\label{fig:phasediag}
\end{figure}

Finally, we turn to the intriguing question of the magnetic behavior of the solid solutions (U$_{1-x}$Np$_{x}$)O$_{2}$ that was reported some years ago, as well as some recent measurements on single crystals at ITU, Karlsruhe. We show in Fig. 4 the experimental points determining the ``ordering'' temperatures as a function of the Np concentration, $x$. Since the ordering interactions in both UO$_{2}$ ($x = 0$) and NpO$_{2}$ ($x = 1$) lead to $T_{0} \sim 25$~K the sudden drop in $T_{0}$ on dilution at either end is surprising, as we would naively expect a Vegard-type law for $T_{0}$. Moreover, for the region  $\sim0.3 < x <  \sim0.8$, neutron and M\"ossbauer experiments show a surprising short-range ordering\cite{tabuteau357,tabuteau373,boeuf221} with propagation wavevector $\langle\frac{1}{2}\frac{1}{2}\frac{1}{2}\rangle$ that is not yet well characterized or understood.

The experiments reported in this paper, as well as earlier work \cite{paixao187202,caciuffos2287,wilkins214402}, show that the important parameter across this phase diagram for all $x$ is antiferroquadrupolar  (AFQ) ordering of the electric quadrupoles. For small $x$, as shown in Fig. 3, the AFQ ordering is transverse, whereas for large $x$ near NpO$_{2}$\cite{paixao187202,caciuffos2287} the ordering is longitudinal.  In both cases the same wave vector $\vec{q}_{AFQ} = \langle001\rangle$ is observed. Our suggestion, therefore, is that over the large intermediate range of $x$ (shown by the shaded region  in Fig.~\ref{fig:phasediag}) a state of frustrated quadrupolar ordering exists, where neither the AFQ-T nor AFQ-L order can dominate, leading to short-range ordering of the quadrupoles and associated dipole moments. As electric quadrupoles are a secondary order parameter in these compounds, the frustration could reflect a competition between Np-Np octupolar interactions favoring longitudinal AFQ and U-Np and U-U dipole interactions favoring transverse AFQ. Further experiments in this interesting region are planned with RXS and other techniques such as NMR.

\acknowledgments
SBW would like to acknowledge support from the European Commission in the framework of the ``Training and Mobility of Researchers'' program. GHL thanks Russ Walstedt of JAERI (Japan) for many interesting discussions about actinide oxides.
\bibliography{../../../bibTeX/UO2-NpO2,../../../bibTeX/xray}
\end{document}